\documentclass[aps,prx,twocolumn,superscriptaddress,longbibliography,floatfix]{revtex4-2}

\usepackage{graphicx}
\usepackage{amsmath,amssymb}
\usepackage{bm}
\usepackage{xcolor}
\usepackage{ulem}
\usepackage{capt-of}

\makeatletter
\newenvironment{widefigureblock}{%
  \par\onecolumngrid\vskip6pt\noindent\ignorespaces
}{%
  \par\vskip6pt\twocolumngrid\global\@ignoretrue\@endpetrue
}
\makeatother

\begin{document}
\raggedbottom

\title{Learning Boson Star Solution Families with Physics-Informed Neural Networks}

\author{Ao Liu}
\altaffiliation{These authors contributed equally to this work.}
\affiliation{Department of Physics, Key Laboratory of Low Dimensional Quantum Structures and Quantum Control of Ministry of Education, Institute of Interdisciplinary Studies, Hunan Research Center of the Basic Discipline for Quantum Effects and Quantum Technologies, and Synergetic Innovation Center for Quantum Effects and Applications, Hunan Normal University, Changsha, Hunan 410081, P. R. China}
\affiliation{Hunan Research Center of the Basic Discipline for Quantum Effects and Quantum Technologies, Hunan Normal University, Changsha 410081, China}
\affiliation{College of Information Science and Engineering, Hunan Normal University, Changsha 410081, China}

\author{Chen-Hao Hao}
\altaffiliation{These authors contributed equally to this work.}
\affiliation{Department of Physics, Key Laboratory of Low Dimensional Quantum Structures and Quantum Control of Ministry of Education, Institute of Interdisciplinary Studies, Hunan Research Center of the Basic Discipline for Quantum Effects and Quantum Technologies, and Synergetic Innovation Center for Quantum Effects and Applications, Hunan Normal University, Changsha, Hunan 410081, P. R. China}
\affiliation{Hunan Research Center of the Basic Discipline for Quantum Effects and Quantum Technologies, Hunan Normal University, Changsha 410081, China}

\author{Cuihong Wen}
\email{cuihongwen@hunnu.edu.cn}
\affiliation{College of Information Science and Engineering, Hunan Normal University, Changsha 410081, China}

\author{Shao-Jiang Wang}
\email{schwang@itp.ac.cn}
\affiliation{Institute of Theoretical Physics, Chinese Academy of Sciences (CAS), Beijing 100190, China}

\author{Jieci Wang}
\email{jcwang@hunnu.edu.cn}
\affiliation{Department of Physics, Key Laboratory of Low Dimensional Quantum Structures and Quantum Control of Ministry of Education, Institute of Interdisciplinary Studies, Hunan Research Center of the Basic Discipline for Quantum Effects and Quantum Technologies, and Synergetic Innovation Center for Quantum Effects and Applications, Hunan Normal University, Changsha, Hunan 410081, P. R. China}
\affiliation{Hunan Research Center of the Basic Discipline for Quantum Effects and Quantum Technologies, Hunan Normal University, Changsha 410081, China}

\date{\today}

\begin{abstract}
Computing boson star families traditionally requires repeated solution of nonlinear eigenvalue boundary-value problems and careful numerical continuation through turning points. We develop a physics-informed neural network (PINN) that learns the map from the physical parameters and radial coordinate directly to the scalar and metric fields over an equilibrium solution manifold. Regularity and asymptotic boundary conditions are incorporated into the network output, while the training objective combines pointwise supervision, Einstein-Klein-Gordon residuals, and curve-level constraints on the Arnowitt-Deser-Misner mass and Noether charge. A trained model generates a complete configuration in a single forward pass. Across representative one-, two-, and three-branch families, the method reconstructs the mass-frequency spirals and conserved quantities, including configurations on inner branches that require delicate continuation in conventional solvers. These results establish physics-informed surrogate learning as a practical route to amortized exploration of nonlinear self-gravitating solution families.
\end{abstract}

\maketitle

\section{Introduction}
\label{sec:introduction}

Boson stars are regular, horizonless compact solutions of self-gravitating bosonic systems, whose origins trace back to pioneering work on gravitationally bound scalar fields~\cite{Kaup:1968zz,Ruffini:1969qy}. In contrast to ordinary stars, they are sustained by coherent scalar or vector fields in curved spacetime~\cite{Colpi:1986ye,Schunck:2003kk,Liebling:2012fv}. Their masses, radii, and compactness can vary substantially with the boson mass and the self-interaction potential. They therefore provide a clean laboratory for nonlinear matter in strong gravity and have been explored in connection with dark matter~\cite{Sahni:1999qe,Hu:2000ke,Hao:2025gak} and black-hole mimickers~\cite{Torres:2000dw,Guzman:2009zz,Vincent:2015xta,Olivares:2018abq,Cardoso:2019rvt}.

The central field amplitude and field frequency serve as natural parameters along an equilibrium family, with the latter acting as a nonlinear eigenvalue. Complete numerical sequences exhibit characteristic spirals in the mass-frequency plane~\cite{Brito:2015pxa,Herdeiro:2017fhv,Liang:2022mjo}. Their multiple branches encode nonlinear gravitational behavior and provide the backgrounds required for systematic studies of stability~\cite{Gleiser:1988ih,Balakrishna:1997ej,Seidel:1990jh,Sanchis-Gual:2019ljs,Kain:2021rmk,Brito:2023fwr}, critical phenomena~\cite{Hawley:2000dt,Jimenez-Vazquez:2022fix,Ma:2024olw,Zhang:2023qxf}, and observational signatures~\cite{Cardoso:2016oxy,Palenzuela:2017kcg,Cunha:2017wao,CalderonBustillo:2020fyi,Evstafyeva:2024qvp,He:2025qmq}.

Boson star configurations are conventionally formulated as nonlinear eigenvalue boundary-value problems. Standard approaches include shooting algorithms~\cite{Herdeiro:2020kvf}, finite-difference or finite-element relaxation~\cite{Kleihaus:2005me,Gervalle:2022fze}, and spectral methods~\cite{Grandclement:2009ju}. These techniques can produce highly accurate individual solutions and global sequences, but shooting is sensitive to initial data and eigenfrequency tuning, while relaxation and spectral methods require tailored grids, initial guesses, and continuation strategies. Parameter changes generally require new scans. Several turning points make the construction still more expensive because each branch must be followed through repeated nonlinear solves.

Physics-informed neural networks (PINNs) offer a complementary strategy~\cite{2021NatRP...3..422K}. Since their introduction for forward and inverse differential-equation problems~\cite{2019JCoPh.378..686R}, PINNs have been applied to fluid mechanics, quantum mechanics, and gravitational and cosmological systems~\cite{2020Sci...367.1026R,2024PhRvL.132a0801N,2022PhRvD.106l4047C,2025PhRvR...7a3164Z,2023PhRvD.107f3523C,Luna:2022rql,Luna:2024spo,Stefanou:2023jxk,Zhou:2026haz,SchettiniGherardini:2026bdb}. Automatic differentiation enforces the governing equations throughout the computational domain, while neural operators and related architectures provide routes to learning parameterized solution maps~\cite{Lu:2021,Li:2021}.

Here we construct an amortized PINN solver for spherical boson star families. Instead of solving the Einstein-Klein-Gordon system independently for every parameter choice, the network learns the entire conditional map from frequency, central amplitude, and radius to the scalar and metric fields. We integrate the reference solver, network construction, training procedure, and tests on one-, two-, and three-branch families into a single account. The resulting surrogate generates full configurations without per-solution nonlinear iteration and remains effective across branch turning points.

\section{Physical Model and Reference Solutions}
\label{sec:model}

\subsection{Einstein-Klein-Gordon system}
\label{sec:ekg}

\begin{figure}[t]
\centering
\includegraphics[width=\columnwidth]{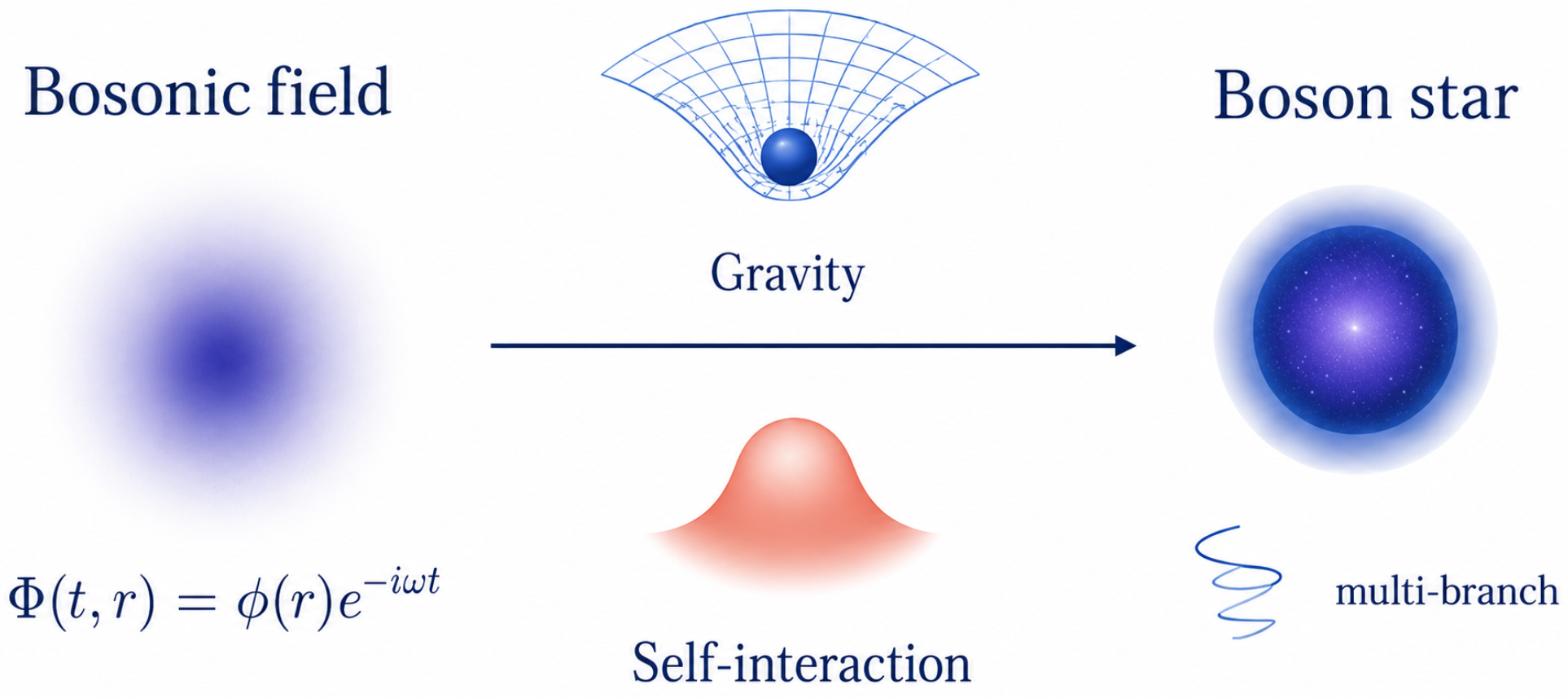}
\caption{Schematic representation of a boson star as a macroscopic coherent bosonic configuration bound by gravity.}
\label{fig:formation}
\end{figure}

We consider a complex scalar field minimally coupled to Einstein gravity, as represented schematically in Fig.~\ref{fig:formation},
with action
\begin{equation}
S=\int \mathrm{d}^4x\sqrt{-g}\left[
\frac{R}{16\pi G}
-g^{\mu\nu}\partial_\mu\Phi^*\partial_\nu\Phi
-U(|\Phi|^2)
\right].
\label{eq:action}
\end{equation}
The quartic-sextic self-interaction potential is
\begin{equation}
U(|\Phi|^2)=\mu^2|\Phi|^2-b|\Phi|^4+c|\Phi|^6,
\label{eq:potential}
\end{equation}
where $b$ and $c$ determine the signs and strengths of the self-interactions. For a static, spherically symmetric configuration we use
\begin{align}
\mathrm{d}s^2&=-n(r)o^2(r)\mathrm{d}t^2+\frac{\mathrm{d}r^2}{n(r)}
+r^2\mathrm{d}\Omega_2^2,\\
\Phi(t,r)&=\phi(r)e^{-i\omega t},
\label{eq:ansatz}
\end{align}
where $\phi(r)$ is real, $\omega$ is the field frequency, and $\kappa\equiv8\pi G$. A prime denotes differentiation with respect to $r$. The two independent Einstein equations are
\begin{align}
n'&=\frac{1-n}{r}
-\kappa r\left(
\frac{\omega^2\phi^2}{no^2}
+n\phi'^2+U
\right),
\label{eq:n_equation}\\
o'&=\kappa ro\left(
\phi'^2+\frac{\omega^2\phi^2}{n^2o^2}
\right),
\label{eq:o_equation}
\end{align}
and the Klein-Gordon equation is
\begin{equation}
\phi''
+\left(
\frac{2}{r}+\frac{n'}{n}+\frac{o'}{o}
\right)\phi'
+\left[
\frac{\omega^2}{n^2o^2}
-\frac{1}{n}\frac{\mathrm{d}U}{\mathrm{d}|\Phi|^2}
\right]\phi=0.
\label{eq:scalar_equation}
\end{equation}

Regular localized solutions satisfy $n(0)=1$, $\phi(0)=\phi_c$, and $\phi'(0)=0$ at the origin, together with $\phi\to0$, $o\to1$, and $n\to1-2GM/r+\mathcal{O}(r^{-2})$ at spatial infinity. The asymptotic metric defines the ADM mass
\begin{equation}
M=\lim_{r\to\infty}\frac{r}{2G}\left[1-n(r)\right].
\label{eq:ADM_mass}
\end{equation}
The global $U(1)$ invariance yields
\begin{equation}
J^\mu=ig^{\mu\nu}
\left(\Phi\,\partial_\nu\Phi^*-\Phi^*\partial_\nu\Phi\right),
\end{equation}
and the associated Noether charge is
\begin{equation}
Q=\int_{\Sigma}\mathrm{d}^3x\,\sqrt{-g}\,J^t.
\label{eq:charge}
\end{equation}
For fixed $(b,c\mu^2)$, each central amplitude selects an eigenfrequency, and the resulting one-parameter family traces a characteristic curve in the $(M,\omega)$ plane.

\subsection{Finite-element Newton solver}
\label{sec:fem}

Reference solutions are generated with a finite-element Newton method on a compactified radial domain,
\begin{equation}
x=\frac{r}{r+L}\in[0,1],
\qquad
r=\frac{Lx}{1-x},
\label{eq:compactification}
\end{equation}
where the endpoint $x=1$ represents spatial infinity. The interval is divided into $N=2000$ grid points, and the unknown functions are expanded in finite-element basis functions,
\begin{align}
n(x)&=\sum_a n_aN_a(x),
&o(x)&=\sum_a o_aN_a(x),\nonumber\\
\phi(x)&=\sum_a\phi_aN_a(x).
\label{eq:fem_expansion}
\end{align}
After compactification, Eqs.~(\ref{eq:n_equation})-(\ref{eq:scalar_equation}) and their boundary conditions are imposed simultaneously. At fixed $\phi_c$, the frequency $\omega$ is an additional nonlinear unknown determined by scalar-field localization.

This solver is accurate and robust for many configurations, but its cost and reliability depend on mesh resolution and the initial guess. Strong self-interaction, inner spiral branches, and large parameter scans require finer grids and controlled continuation steps. These increasingly difficult nonlinear solves motivate learning a reusable conditional solution map from a limited reference set.

\section{Physics-Informed Surrogate Framework}
\label{sec:framework}

\subsection{Multiscale data organization}
\label{sec:data}

The numerical solutions are organized at two complementary levels. A point-level sample treats every radial grid point as
\begin{equation}
(x,\omega,\phi_c)\longrightarrow(\phi,n,o),
\label{eq:point_map}
\end{equation}
and supports direct supervision of the local fields. Curve-level samples retain entire configurations so that $M$ and $Q$ can be reconstructed and compared with their reference values. This separation is important because small pointwise discrepancies can accumulate into appreciable errors in integrated observables. The network operates on compactified $x$, while field-equation residuals and global quantities are evaluated in physical radius $r$.

\subsection{Architecture and constrained output}
\label{sec:architecture}

\begin{figure*}[t]
\centering
\includegraphics[width=\textwidth]{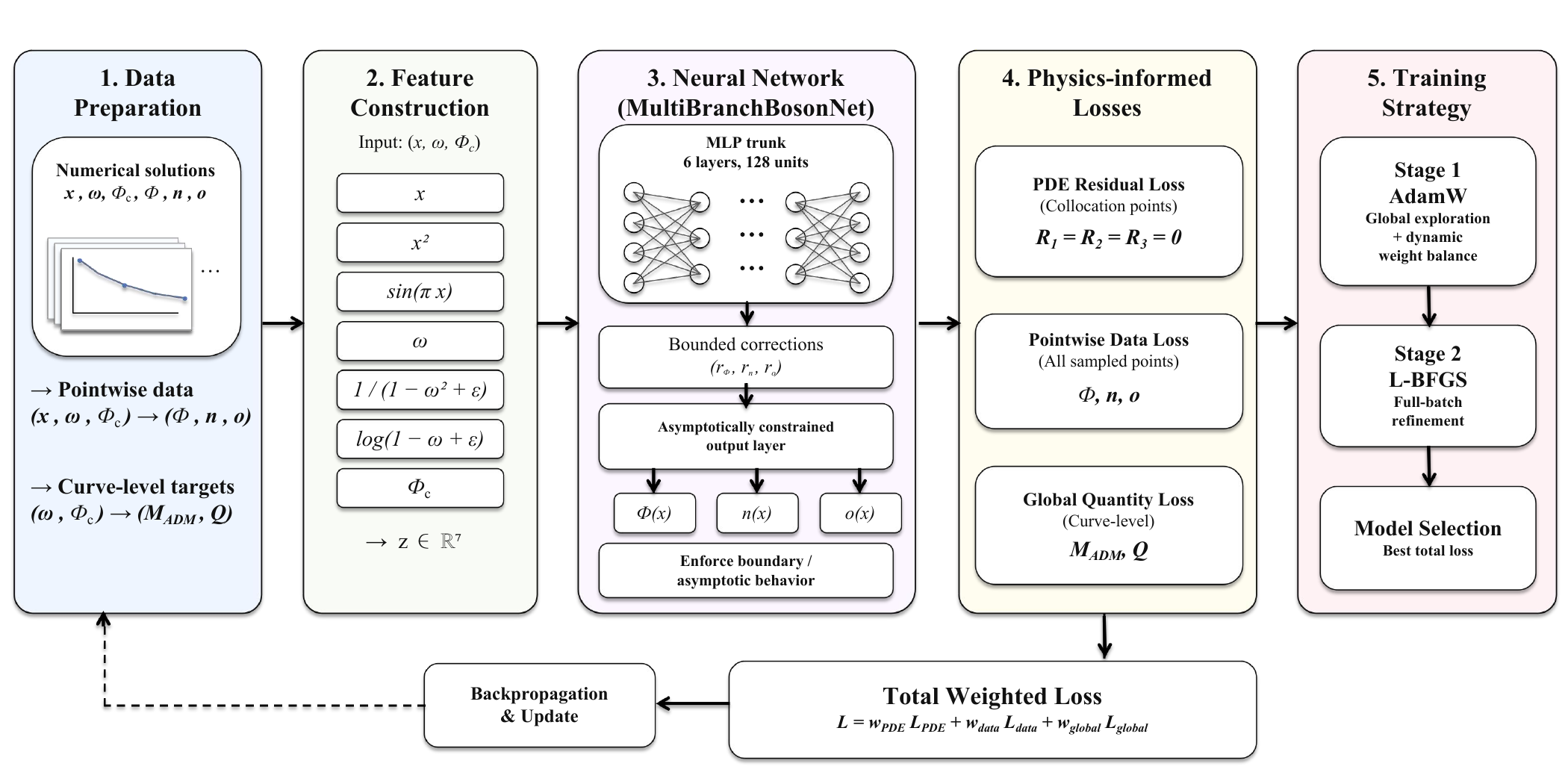}
\caption{Workflow of the physics-informed surrogate. Reference solutions are converted into point- and curve-level samples and augmented by physics-inspired input features. The MLP predicts bounded corrections to an asymptotically constrained output. Training combines field data, differential-equation residuals, and global constraints on $M$ and $Q$.}
\label{fig:workflow}
\end{figure*}

Figure~\ref{fig:workflow} summarizes the framework: physics-inspired feature construction is followed by a fully connected multilayer perceptron (MLP) and a boundary-condition-aware output layer.

For each sampling point, $(x,\omega,\phi_c)$ is mapped to
\begin{equation}
\mathbf{z}=
\left(
x,x^2,\sin(\pi x),\omega,
\frac{1}{1-\omega^2+\varepsilon},
\log(1-\omega+\varepsilon),\phi_c
\right),
\label{eq:features}
\end{equation}
where $\varepsilon=10^{-4}$ regularizes the near-critical transformations. The features $x^2$ and $\sin(\pi x)$ enrich the radial representation, while the rational and logarithmic frequency features resolve the rapidly changing regime $\omega\to\mu$. Because $\omega(\phi_c)$ is not globally monotonic on a folded equilibrium sequence, $\omega$ alone does not uniquely label a configuration. The joint input $(\omega,\phi_c)$ therefore resolves the branch ambiguity and identifies the configuration on the sampled equilibrium manifold.

The feature vector is processed by six hidden layers of 128 neurons with Tanh activations, whose smooth first and second derivatives are suitable for automatic differentiation. The MLP produces unconstrained latent outputs $(\hat r_\phi,\hat r_n,\hat r_o)$, which are compressed to bounded corrections,
\begin{equation}
r_i=S\tanh\left(\frac{\hat r_i}{S}\right),
\qquad i\in\{\phi,n,o\},
\qquad S=5.
\label{eq:bounded_correction}
\end{equation}
With $k=\sqrt{\mu^2-\omega^2}$, the physical outputs are
\begin{align}
\phi(r)&=\phi_c e^{-kr}(1+kr)
+x^2(1-x)e^{-kr}r_\phi,
\label{eq:output_phi}\\
n(r)&=1+x^2(1-x)r_n,\nonumber\\
o(r)&=1+(1-x)^2r_o.
\label{eq:output_metric}
\end{align}
The scalar base function satisfies $\phi(0)=\phi_c$ and $\phi'(0)=0$, while its exponential factor supplies the leading far-field decay. The $x^2$ factors preserve origin regularity, and the factors that vanish at $x=1$ enforce the scalar and metric asymptotics. The network therefore learns only bounded interior corrections instead of rediscovering the boundary behavior from data.

\subsection{Local, physical, and global losses}
\label{sec:loss}

Training minimizes
\begin{equation}
L=w_{\rm data}L_{\rm data}
+w_{\rm PDE}L_{\rm PDE}
+w_{\rm global}L_{\rm global},
\label{eq:total_loss}
\end{equation}
where
\begin{equation}
L_{\rm data}=L_\phi+L_n+L_o
\label{eq:data_loss}
\end{equation}
is the sum of the fieldwise mean-squared errors.

Automatic differentiation gives derivatives with respect to $x$, which are transformed to $r$ by the chain rule. The residual form used in training is
\begin{equation}
\begin{aligned}
R_1={}&r(\omega^2-m_{\rm eff}^2no^2)\phi\\
&+no\left[
\left(
ro\frac{\mathrm{d}n}{\mathrm{d}r}
+n\left(2o+r\frac{\mathrm{d}o}{\mathrm{d}r}\right)
\right)\frac{\mathrm{d}\phi}{\mathrm{d}r}
+rno\frac{\mathrm{d}^2\phi}{\mathrm{d}r^2}
\right],
\end{aligned}
\label{eq:R1}
\end{equation}
with
\begin{equation}
m_{\rm eff}^2=\mu^2-2b\phi^2+3c\phi^4,
\end{equation}
and
\begin{align}
R_2={}&-\frac1r+\frac nr+\frac{\mathrm{d}n}{\mathrm{d}r}\nonumber\\
&+\kappa r\left(
\mu^2+\frac{\omega^2}{no^2}-b\phi^2+c\phi^4
\right)\phi^2\nonumber\\
&+\kappa rn\left(\frac{\mathrm{d}\phi}{\mathrm{d}r}\right)^2,
\label{eq:R2}\\
R_3={}&\frac{\mathrm{d}o}{\mathrm{d}r}
-\kappa ro\left(\frac{\mathrm{d}\phi}{\mathrm{d}r}\right)^2
-\kappa r\frac{\omega^2\phi^2}{n^2o}.
\label{eq:R3}
\end{align}
The physics loss is
\begin{equation}
L_{\rm PDE}=w_{R_1}\langle R_1^2\rangle
+w_{R_2}\langle R_2^2\rangle+w_{R_3}\langle R_3^2\rangle.
\label{eq:pde_loss}
\end{equation}
The adaptive residual weights balance the Klein-Gordon and Einstein equations during training as their relative loss magnitudes evolve.

For each curve-level sample, the complete predicted profile is used in Eqs.~(\ref{eq:ADM_mass}) and (\ref{eq:charge}). The resulting contribution,
\begin{equation}
L_{\rm global}
=w_M\left(M^{\rm pred}-M^{\rm true}\right)^2
+w_Q\left(Q^{\rm pred}-Q^{\rm true}\right)^2,
\label{eq:global_loss}
\end{equation}
links local accuracy to integral consistency. Thus boundary constraints, differential residuals, and global observables act at complementary physical scales.

\section{Training Strategy}
\label{sec:training}

Optimization proceeds in two stages. AdamW is used for $10{,}000$ epochs with a base learning rate of $5\times10^{-4}$ and weight decay $10^{-6}$. A 300-epoch linear warmup raises the learning rate from $0.1$ to $1$ times its base value, after which cosine annealing with warm restarts uses $T_0=2500$ and $\eta_{\min}=10^{-6}$. Each iteration samples $40{,}000$ collocation points in $x\in(0,0.99)$, and gradient norms are clipped at $0.2$. A subsequent L-BFGS refinement runs for 500 iterations with strong-Wolfe line search and history size 60 on a batch fixed at the start of this full-batch phase. Table~\ref{tab:hyperparams} summarizes the setup.

\begin{table}[t]
\centering
\caption{Training hyperparameters. Adaptive weights are listed at their initial values.}
\label{tab:hyperparams}
\begin{tabular}{ll}
\hline
Parameter & Value\\
\hline
AdamW epochs & $10{,}000$\\
L-BFGS iterations & $500$\\
Base learning rate & $5\times10^{-4}$\\
Learning-rate warmup & $300$ epochs\\
Cosine restart period $T_0$ & $2500$\\
Weight decay & $10^{-6}$\\
Gradient clipping norm & $0.2$\\
Collocation batch size & $40{,}000$\\
L-BFGS history size & $60$\\
Residual weights $w_{R_i}$ (initial) & $1.0$\\
Data weight $w_{\rm data}$ (initial) & $25.0$\\
PDE weight $w_{\rm PDE}$  & $1.0$\\
Data weight $w_{\rm global}$  & $1.0$\\
Adaptive-weight rate $\alpha_{\rm w}$ & $0.1$\\
Global mass weight $w_M$ & $6.0$\\
Global charge weight $w_Q$ & $0.1$\\
Curve-level warmup & $1500$ epochs\\
Curve batch size & $64$\\
\hline
\end{tabular}
\end{table}

Rather than fixing all relative weights, $w_{R_1}$, $w_{R_2}$, $w_{R_3}$ and $w_{\rm data}$ are updated every five epochs using
\begin{equation}
w_i\leftarrow(1-\alpha_{\rm w})w_i+\alpha_{\rm w}\widetilde w_i,
\qquad \alpha_{\rm w}=0.1,
\label{eq:adaptive_weight}
\end{equation}
where $\widetilde w_i$ reflects the current relative raw-loss magnitudes. This moving average prevents the stiff near-critical Klein-Gordon residual from either dominating or being overwhelmed by the metric and data terms across different $(b,c\mu^2)$ models.

Global supervision is introduced gradually because imposing integral constraints before the local fields are accurate can destabilize training. Its warmup factor is
\begin{equation}
s(e)=\min\left(1,\frac{e}{1500}\right),
\label{eq:global_warmup}
\end{equation}
where $e$ is the epoch. Once activated, the mass and charge weights are $w_M=6.0$ and $w_Q=0.1$. Each step samples 64 curves, with configurations satisfying $\omega>0.95\mu$ assigned three times the sampling probability of the remaining data.

\section{Results}
\label{sec:results}

We set $\mu=1$ and $\kappa=2$ and study three qualitatively different models: a one-branch family with $(b,c\mu^2)=(150,4000)$, a two-branch family with $(50,2000)$, and a three-branch family with $(-20,0)$. A separate network is trained for each model using $20\%$ of the reference configurations and evaluated on the remaining unseen configurations. Relative errors are reported for the nonzero global observables $M$ and $Q$. Absolute errors are used for local fields because the pointwise relative scalar error becomes ill-conditioned as $\phi_{\rm ref}\to0$; for the order-unity metric functions, absolute and relative maps have essentially the same structure.

Two physical relations guide the interpretation. First, in the asymptotically flat
region the scalar field behaves as $\phi(r)\sim A e^{-kr}/r$, with $k=\sqrt{\mu^2-\omega^2}$, so its exponential localization scale $\ell_{\rm tail}\sim k^{-1}$ grows rapidly as $\omega\to\mu$. Second, neighboring equilibrium configurations satisfy the boson-star first-law relation $\mathrm{d}M=\omega\,\mathrm{d}Q$; correlated structure in the predicted $M(\omega)$ and $Q(\omega)$ curves is therefore a physical consistency test rather than two independent regression checks. Turning points of these curves are also natural candidates for changes in radial stability under standard turning-point arguments~\cite{Gleiser:1988ih,Kain:2021rmk,Brito:2023fwr}, although establishing stability requires a separate perturbation analysis.

\subsection{One-branch family}
\label{sec:one_branch}

\begin{figure}[t]
\centering
\includegraphics[width=\columnwidth]{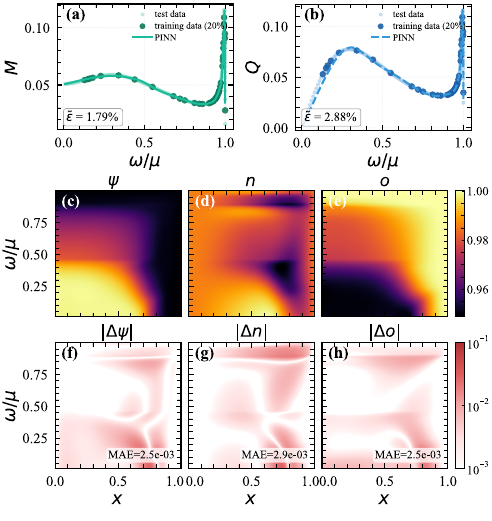}
\caption{Reconstruction of the one-branch family for $(b,c\mu^2)=(150,4000)$. Panels (a) and (b) show $M$ and $Q$ versus $\omega/\mu$; dark circles are training configurations, light circles are test configurations, and solid curves are predictions. Panels (c)-(e) show predicted $\phi$, $n$, and $o$, while (f)-(h) show their absolute errors.}
\label{fig:single_branch}
\end{figure}

Figure~\ref{fig:single_branch} summarizes the results for the one-branch model. In panels (a) and (b), the predicted $M(\omega)$ and $Q(\omega)$ curves closely follow both the training and unseen test configurations over the full frequency range. The mean relative errors are $1.79\%$ for $M$ and $2.88\%$ for $Q$. The locations and overall shapes of the maxima are also reproduced without visible discontinuities.

Panels (c)-(e) show that the network reconstructs the main variations of $\phi$, $n$, and $o$ across both the radial coordinate and the frequency. The corresponding error maps in panels (f)--(h) are predominantly at the $10^{-3}$ level, with narrow regions of larger error near rapidly varying parts of the data. The mean absolute errors are $2.5\times10^{-3}$, $2.9\times10^{-3}$, and $2.5\times10^{-3}$ for $\phi$, $n$, and $o$, respectively.

\subsection{Two-branch family}
\label{sec:two_branch}

Figure~\ref{fig:two_branch} compares the predictions with the reference data for both branches. The predicted $M(\omega)$ and $Q(\omega)$ curves pass through the training and test configurations and remain continuous around the turning point. For Branches I and II, the mean relative errors are $0.85\%$ and $0.27\%$ for $M$, and $0.60\%$ and $0.20\%$ for $Q$, respectively. The maximum-mass region and the short second branch are both recovered accurately.

The field maps for both branches reproduce the principal variations of $\phi$, $n$, and $o$. Their mean absolute errors range from $6.2\times10^{-3}$ to $2.8\times10^{-2}$. The largest deviations appear in relatively narrow bands, mainly toward larger
$x$, while the absolute errors remain of order $10^{-2}$ or smaller over most of the domain. These results indicate that the model can distinguish the two branches and reconstruct their local fields as well as their global quantities.

\begin{widefigureblock}
\begin{minipage}{\textwidth}
\centering
\includegraphics[width=0.84\textwidth]{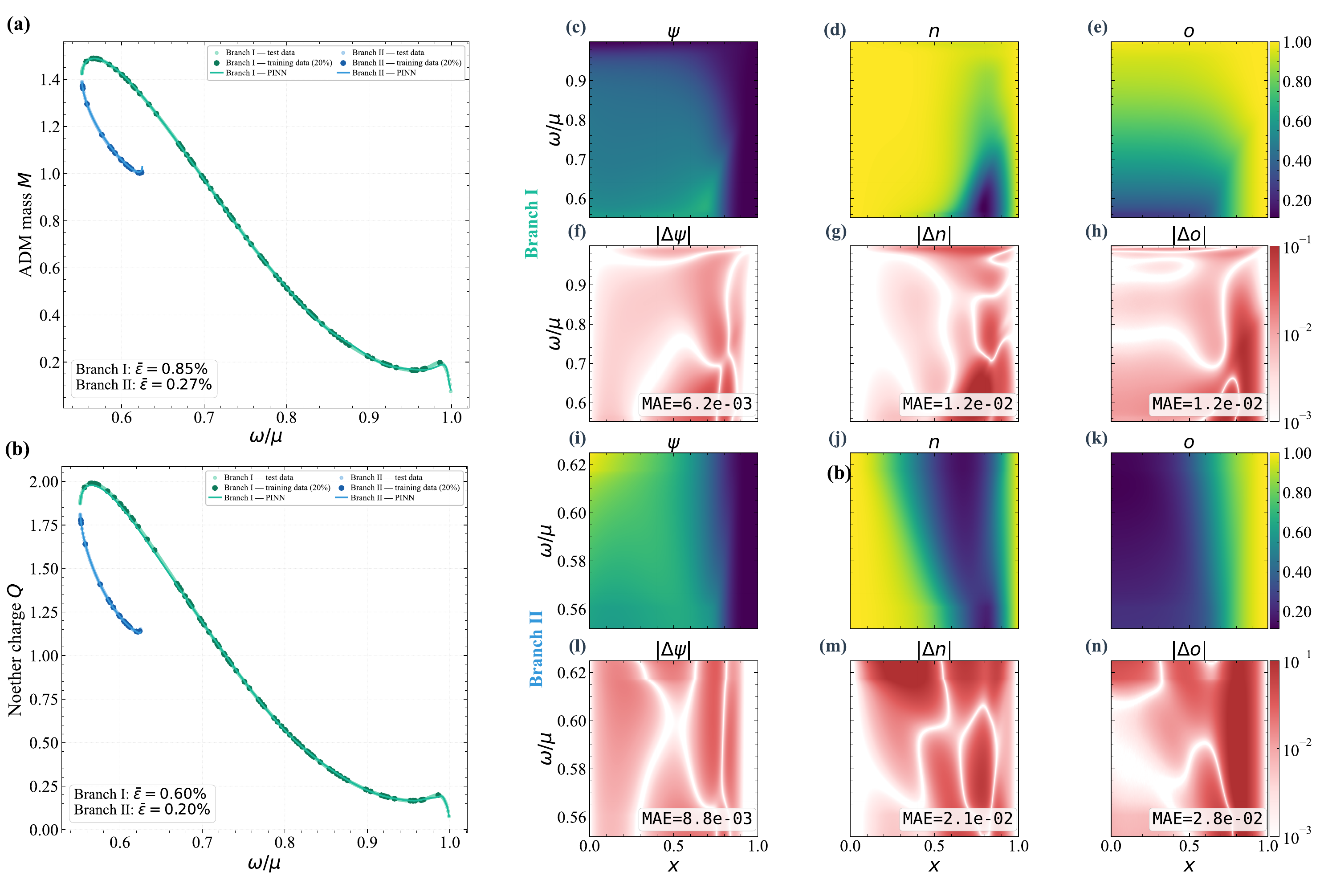}
\captionof{figure}{Reconstruction of the two-branch family for $(b,c\mu^2)=(50,2000)$. Green and blue denote Branches I and II. Panels (c)-(h) and (i)-(n) show reconstructed fields and absolute errors for the two branches; panels of the same field share a common color scale. Symbols and curves follow Fig.~\ref{fig:single_branch}.}
\label{fig:two_branch}
\end{minipage}
\end{widefigureblock}

\subsection{Three-branch family}
\label{sec:three_branch}

Figure~\ref{fig:three_branch} presents the most complicated case. The predicted mass and charge curves reproduce all three branches and the two turning regions. On Branches I and II, the mean relative errors are $0.68\%$ and $0.62\%$ for $M$, and $0.48\%$ and $0.41\%$ for $Q$. On Branch III, the errors increase to $2.63\%$ for $M$ and $0.76\%$ for $Q$. The agreement is therefore strongest on the long outer branch and becomes less accurate on the short innermost branch.

The reconstructed field maps follow the reference trends on all three branches. On Branch I, most local errors lie between $\mathcal{O}(10^{-3})$ and $\mathcal{O}(10^{-2})$. The errors increase to approximately $\mathcal{O}(10^{-2})$ on Branches II and III, and the largest value is $|\Delta n|=6.4\times10^{-2}$ on Branch III. Nevertheless, the overall patterns of the three fields and the separation between the branches remain clearly reproduced.

The relatively larger errors occur in the short inner branches, near the turning regions, and close to the outer radial boundary. Possible reasons include the smaller number of training configurations in these regions, the faster variation of the target fields, and the imbalance between easy and difficult samples during optimization. Further improvements may be obtained by increasing the sampling density near the turning points and at large $x$, using branch-aware or error-adaptive loss weights, increasing the network capacity, and performing a short branch-specific fine-tuning stage after the joint training.

\begin{widefigureblock}
\begin{minipage}{\textwidth}
\centering
\includegraphics[width=0.72\textwidth,height=0.56\textheight,keepaspectratio]{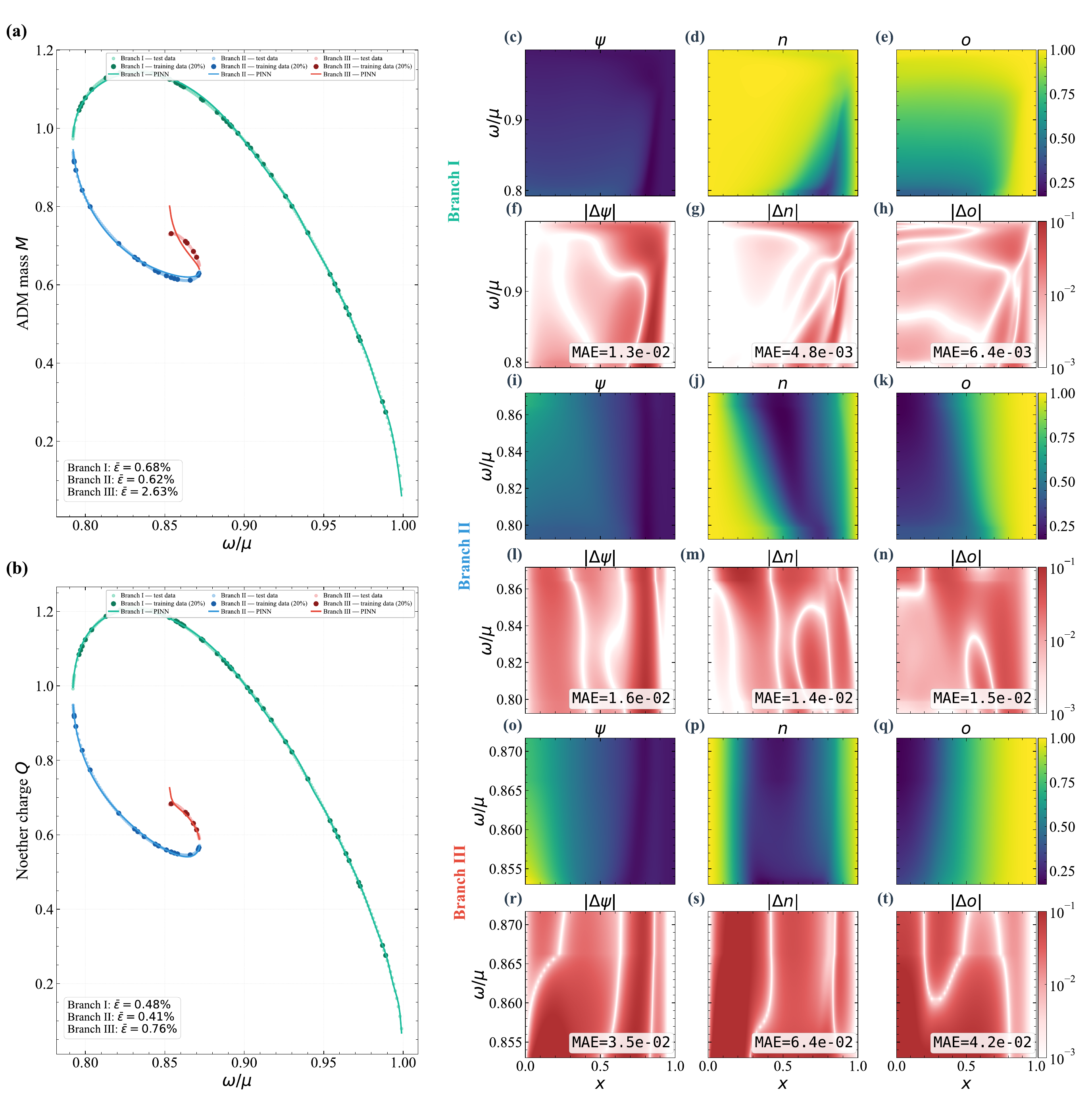}
\captionof{figure}{Reconstruction of the three-branch family for $(b,c\mu^2)=(-20,0)$. Green, blue, and red denote Branches I, II, and III. Panels (c)-(h), (i)-(n), and (o)-(t) show the fields and absolute errors for the three branches. Symbols and curves follow Fig.~\ref{fig:single_branch}.}
\label{fig:three_branch}
\end{minipage}
\end{widefigureblock}

\section{Discussion}
\label{sec:discussion}

The principal advantage of the surrogate is amortization. A conventional solver must retune an eigenfrequency and perform nonlinear iterations for each configuration, while a trained PINN evaluates an entire profile in one forward pass. At a turning point, the learned conditional map produces a smooth sequence without changing continuation parameters or repeatedly modifying initial data. This feature is valuable when dense background families are needed for stability, catastrophe-theory, critical-phenomena, or observational calculations.

The branch structure reconstructed here is not merely a geometric feature of the
$M$-$\omega$ diagrams, but reflects the nonlinear equilibrium of the
self-gravitating scalar field. For fixed self-interaction parameters, the solutions
form a one-parameter family that may be parametrized by the central amplitude
$\phi_c$, while $\omega=\omega(\phi_c)$ is determined as a nonlinear eigenvalue.
In the relativistic regime, this relation can become nonmonotonic and fold back on
itself, so that its projection onto the $M$-$\omega$ plane becomes multivalued:
the same frequency can correspond to physically distinct stars with different
central amplitudes and radial profiles. Changing the scalar potential modifies the
nonlinear restoring term $m_{\rm eff}^2=\mu^2-2b\phi^2+3c\phi^4$, and hence
reshapes the balance among self-interaction, gradient support, and gravitational
binding, producing the different one-, two-, and three-branch structures sampled
here. From this viewpoint, these three cases represent different fold geometries of
the underlying equilibrium manifold rather than unrelated regression tasks. This
also clarifies the learning behavior: near a fold,
$\mathrm{d}\omega/\mathrm{d}\phi_c$ becomes small or changes sign, making
$\omega$ alone a locally ill-conditioned label for the solution. Supplying
$\phi_c$ together with $\omega$ resolves this branch ambiguity and allows the
network to learn a single-valued representation of the folded manifold, while the
larger errors on the short inner branches are consistent with their stronger local
variation and sparser sampling~\cite{Schunck:2003kk,Liebling:2012fv}.

The surrogate is not a replacement for high-precision finite-element or spectral solvers. Such solvers remain necessary to generate reference data and to certify individual configurations. Instead, the PINN converts a limited set of expensive solutions into a reusable representation of a family. The results also expose the present limitations: accuracy degrades on compact, sparsely sampled inner branches; each interaction model is currently assigned a separate network; and the present training domain does not include $b$ and $c\mu^2$ as conditional inputs.

Denser branch-aware sampling, branch-specific weighting, and increased model capacity may improve the innermost configurations. A more ambitious extension is to condition the network on the interaction parameters themselves, creating a single surrogate across potentials. The same construction could then be adapted to other nonlinear gravitational systems, including Proca stars, wormholes, and neutron stars, provided suitable boundary-aware outputs and global constraints are identified.

\section{Conclusion}
\label{sec:conclusion}

We have developed a physics-informed surrogate for complete spherical boson star families. The method combines an asymptotically constrained output parameterization with local field supervision, Einstein--Klein--Gordon residuals, and global mass and charge constraints. It reconstructs representative one-, two-, and three-branch solution manifolds and remains effective at the turning points that make conventional continuation costly. The approach therefore offers a practical route to rapid parameter surveys while retaining direct connections to the governing equations and conserved quantities. Extending the learned domain to interaction parameters and improving the resolution of compact inner branches are natural next steps.

\begin{acknowledgments}
This work was supported by the National Natural Science Foundation of China (Grants No.~12374408, No.~12475051, No.~12547147, No.~12422502, No.~12547110, No.~12588101, No.~12235019, and No.~12447101), and the National Key Research and Development Program of China (Grants No.~2021YFC2203004 and No.~2021YFA0718304).
\end{acknowledgments}

\end{document}